\documentclass[twocolumn,aps,prl,superscriptaddress]{revtex4}
\usepackage[utf8]{inputenc}
\usepackage{epsf,graphicx}
\usepackage{multirow}
\usepackage{amsmath,gensymb,amssymb}
\usepackage{lipsum}
\usepackage{dcolumn}
\usepackage{bm}
\usepackage{hyperref}
\usepackage{natbib}
\usepackage{color}

\begin{document}

	\title{Simulation of the Wave Turbulence of a Liquid Surface Using the Dynamic Conformal Transformation Method}

	\author{E.A. Kochurin}
	\email{kochurin@iep.uran.ru}
	\affiliation{Institute of Electrophysics, Ural Division, Russian Academy of Sciences, Yekaterinburg, 620016 Russia}
	\affiliation{Skolkovo Institute of Science and Technology, 121205, Moscow, Russia}

	\begin{abstract}
		The dynamic conformal transformation method has been generalized for the first time to numerically simulate the capillary wave turbulence of a liquid surface in the plane symmetric anisotropic geometry. The model is strongly nonlinear and involves effects of surface tension, as well as energy dissipation and pumping. Simulation results have shown that the system of nonlinear capillary waves can pass to the quasistationary chaotic motion regime (wave turbulence). The calculated exponents of spectra do not coincide with those for the classical Zakharov–Filonenko spectrum for isotropic capillary turbulence but are in good agreement with the estimate obtained under the assumption of the dominant effect of five-wave resonant interactions.
	\end{abstract}
	
	\maketitle


{\bf Introduction.}
It is well known that nonlinear wave systems can pass to a quasistationary chaotic state (wave turbulence regime) due to resonant wave interactions\cite{zakh}. Wave turbulence can occur in an arbitrary nonlinear wave system. In particular, optical turbulence \cite{pic2014}, magnetic and electrohydrodynamic wave turbulence \cite{gal18,gal2000,koch22,Falcon11,koch23}, acoustic turbulence \cite{ZS70,Naz22,JETPL22} and the turbulence of dispersive capillary and gravity waves on the free surface of a liquid \cite{ZF67,korot23,Pan22} are known.  The authors of \cite{ZF67, ZS70} obtained for the first time exact solutions of the
kinetic equations for the distribution function of quasiparticle waves that describe the steady-state energy transfer (or other integrals of motion) over various scales. Such solutions were called Kolmogorov-Zakharov turbulence spectra by analogy with classical hydrodynamic turbulence\cite{zakh}.

The spectrum of isotropic capillary turbulence on a liquid surface (also known as the Zakharov-Filonenko \cite{ZF67}) has been already confirmed with a
high accuracy both experimentally \cite{kolmakov, Falcon22} and numerically \cite{push97, Falcon14, pan14}. The Zakharov-Filonenko spectrum is usually represented in terms of the Fourier spectra of the function $\eta({\bf r},t)$ specifying the shape of the liquid surface as $S_{\eta}(k)=|\eta_\textbf{k}|^2$ and $S_{\eta}(\omega )=|\eta_\omega|^2$:
$$ S_{\eta}(\omega)=C_{3w}^{\omega} P^{1/2}(\sigma/\rho)^{1/6}\omega^{-17/6},$$
\begin{equation}\label{ZF}
S_{\eta}(k)=C_{3w}^k P^{1/2}(\sigma/\rho)^{-3/4}k^{-15/4},\quad k=|{\bf k}|,
\end{equation}
where $\textbf{k}$ is the wave vector, $\omega$ is the angular frequency, $C_{3w}^k$ and $C_{3w}^{\omega}$ are the dimensionless constants, $P$ is the
energy dissipation rate per unit surface area, $\sigma$ and $\rho$ are the surface tension and mass density of the liquid, respectively. The spatial and frequency spectra given by Eqs. (\ref{ZF}) are related to each other by the energy conservation law in the Fourier space; i.e., $S_{\eta}(k)d\textbf{k}=S_{\eta}( \omega)d\omega$. The power-law dependences on $P$ with an exponent of 1/2 in spectra (1) reflect a resonant character
of three-wave interactions:
\begin{equation}\label{3w}
\omega=\omega_1+\omega_2,\qquad \textbf{k}=\textbf{k}_1+\textbf{k}_2,
\end{equation}
The frequency $\omega$ is related to the wavenumber $k$ by the dispersion relation $\omega=(\sigma/\rho)^{1/2}k^{3/2}$.

Thus, Eq. (\ref{ZF}) undoubtedly reproduces the isotropic capillary turbulence spectrum. The situation is different in the case of anisotropic perturbations of the
surface where the considered waves propagate in the same direction, i.e., are collinear. In this case, conditions (\ref{3w}) for the thee-wave resonant interaction are no longer satisfied. The turbulence of collinear capillary waves was numerically studied for the first time in \cite{koch2020} with a weakly nonlinear model. Only trivial three- and four-wave resonant interactions were revealed. Trivial wave resonances do not lead to the energy transfer over scales. Thus, the mechanism of the development of capillary turbulence in the plane symmetric geometry is still unclear. The experimental study carried out in \cite{Ricard21} for collinear waves on the surface of mercury shows the dominant effect of five-wave resonant interactions corresponding to fourth-order nonlinearity. Wave interaction conditions can always be satisfied for resonances of such a high order. Applying the dimensional analysis of weak turbulence spectra (see \cite{naz11}) the authors of \cite{Ricard21} proposed the following estimates for the spectrum of capillary wave turbulence in the quasi-one-dimensional geometry:
$$
S_{\eta}(\omega)=C_{5w}^{\omega} P^{1/4}(\sigma/\rho)^{5/12}\omega^{-31/12},
$$
\begin{equation}\label{5w}
S_{\eta}(k)=C_{5w}^k P^{1/4}(\sigma/\rho)^{-3/8}k^{-27/8},
\end{equation}
where $C_{5w}^k$ and $C_{5w}^\omega$ are the corresponding dimensionless constants. It is noteworthy that the authors of \cite{five} obtained similar turbulence spectra for plane gravity waves. The complete theory of weak turbulence of collinear capillary waves has not yet been proposed. The experimental results obtained in \cite{Ricard21} are in good agreement with analytical spectra (\ref{5w}). The aim of this work is to study the possibility of obtaining spectra (\ref{5w}) in a direct numerical simulation of the turbulence of capillary waves.

When studying plane symmetric capillary turbulence, it is the most difficult to take into account high order nonlinear effects. The standard approach to the study of wave turbulence is based on weakly nonlinear equations of motion \cite{korot16,new11,tran23}. To correctly describe the five-wave interactions, it is necessary to exactly solve a system of equations with fourth-order nonlinearity. The numerical simulation of this system is a rather complicated problem. For this reason, the completely nonlinear approach is used in this work. The analysis is based on the nonstationary conformal transformation method with the transformation of the region filled with the liquid to a half-space, see \cite{ovs74,dya96, dya2002,tan1,tan2,tan3}. The fundamental advantage of this method is the reduction of the initial spatially two-dimensional problem to a one-dimensional system of equations, which directly describes the motion of the liquid surface. The conformal transformation method appeared very convenient to describe nonlinear waves on free surfaces of liquids, see, e.g., \cite{ruban20,dya16,korot19,Nachbin,gao19,gao22,koch18,paras1,paras2}. It is worth noting that the dynamic conformal transformation method has not yet been used to describe the wave turbulence of the free surface of the liquid.

{\bf Model equations.}
The strongly nonlinear dynamics of an ideal incompressible deep liquid with a free surface is considered under the assumption that the motion of the
liquid is plane symmetric; i.e., the full physical model is two-dimensional. Let the Cartesian coordinate system $\{x,y\}$ be such that the equation $y=\eta(x,t)$ determines the deviation of the free surface from the unperturbed state $y=0$. The velocity potential of the liquid $\phi(x,y,t)$ satisfies the Laplace equation:
\begin{equation}\nonumber
\Delta \phi=0.
\end{equation}
At the liquid surface $y=\eta(x,t)$, the dynamic and kinematic boundary conditions are imposed in the form:
\begin{equation}\label{eqx1}
\phi_t+\frac{1}{2}|\nabla \phi|^2=-g \eta +\frac{\sigma}{\rho}\frac{\eta_{xx}}{(1+\eta_x^2)^{3/2}},
\end{equation}
\begin{equation}\label{eqx2}
\eta_t+\eta_x\phi_x=\phi_y,
\end{equation}
where $\nabla=\{\partial_x, \partial_y\}$ and $g$ is the gravitational acceleration. The motion of the liquid vanishes with increasing depth; i.e., $\phi\to 0$ at $y\to -\infty$. The total energy of the system (Hamiltonian) has the form
$$
H=\frac{1}{2}\iint \limits_{y\leq \eta}{|\nabla \phi|^2}dxdy$$
\begin{equation}\label{Ham}+\int\limits_{-\infty}^{+\infty}\left[\frac{g}{2}\eta^2+\frac{\sigma}{\rho}\left(\sqrt{1+\eta_x^2}-1\right)\right]dx.
\end{equation}

The system of Eqs. (\ref{eqx1}) and (\ref{eqx2}) can be represented in terms of the variational derivatives of the Hamiltonian (see \cite{ZF67}):
\begin{equation} \nonumber
\eta_t=\delta H/\delta \psi,\qquad \psi_t=-\delta H/\delta \eta,
\end{equation}
where the functions $\eta(x,t)$ and $\psi=\phi(x,y=\eta,t)$ are the canonically conjugate variables. This system of equations describes the fully nonlinear evolution of capillary–gravity waves on the free surface of the liquid. The minimum phase velocity of linear capillary-gravity waves is reached at the wavelength $\lambda_0=2\pi(\sigma/g \rho)^{1/2}$ corresponding to the period $t_0=2\pi(\sigma/ g^3\rho)^{1/4}$. Below, the dynamics of small-scale capillary
waves with wavenumbers $k\gg 2\pi/\lambda_0$, is considered; i.e., the effect of the gravitational force is neglected. The passage to the dimensionless variables
is performed by setting $\sigma=1$, $\rho=1$. In the linear approximation, the described system is reduced to the dispersion relation:
\begin{equation}\label{disp}
\omega_k=k^{3/2}.
\end{equation}

The region filled with the liquid is then conformally transformed to the half-plane of the new conformal variables $\{u,v\}$. The liquid surface corresponds to
the $v=0$ line:
\begin{equation}\nonumber
y=Y(u,t),\quad \psi=\Psi(u,t),\quad X=u-\hat H Y(u,t).
\end{equation}
Here, $\hat H$ is the Hilbert transform defined in the Fourier space as $\hat H f_k=i\cdot \mbox{sign}(k)f_k$. The profile of the free surface is determined in the parametric form $\eta(x,t)=Y(X(u,t),t)$. Since the procedure of deriving the equations of motion of the liquid in the conformal variables is well known (see \cite{dya96, dya2002,ovs74}), the equations of motion are written below with additional terms describing the energy dissipation and pumping:
\begin{equation}\label{eq1}
Y_t=\left(Y_u\hat H-X_u\right)\frac{\hat H \Psi_u}{J}- \hat \gamma_k Y,
\end{equation}
$$
\Psi_t=\frac{(\hat H \Psi_u)^2-\Psi_u^2}{2J}+\hat H\left(\frac{\hat H \Psi_u}{J}\right)\Psi_u+\frac{X_u Y_{uu}-Y_u X_{uu}}{J^{3/2}}$$
\begin{equation}\label{eq2}
+\mathcal{F}(\textbf{k},t)- \hat \gamma_k \Psi,
\end{equation}
where $J=X_u^2+Y_u^2$ is the Jacobian of the transformation, $\hat \gamma_k$ is the viscosity operator, and $F(\textbf{k},t)$ is the random driving force acting at large scales. The terms responsible for the energy dissipation and pumping are defined in the Fourier space as follows:
\begin{eqnarray*}
\hat \gamma _{k} &=&0,\quad k\leq k_{d}, \\
\hat \gamma _{k} &=&\gamma _{0} (k-k_d)^4,\quad k>k_{d}, \\
\mathcal{F}(\textbf{k},t) &=&F(k)\cdot \exp [iR(\textbf{k},t)], \\
F(k) &=&F_{0}\cdot \exp [-(k-k_{1})^{4}/k_{2}],\quad k\leq k_{2}, \\
F(k) &=&0,\quad k>k_{2}.
\end{eqnarray*}
Here, $R(\mathbf{k,}t)$ are random numbers uniformly distributed in the interval $[0,2\pi ]$, $\gamma _{0}$ and $F_{0}$ are the constant, $k_1$ is the wavenumber at which the pump amplitude reaches the maximum, $k_2$ specifies the width of the pump, and $k_d$ is the scale at which dissipation occurs. The fourth power of $k$ in the viscosity operator was taken by analogy with \cite{korot19}. In terms of the conformal variables, the total energy of the system given by (\ref{Ham}) is represented in the form
\begin{equation}\label{ham2}
H=\frac{1}{2}\int\limits_{-\infty}^{+\infty} \left[ -\Psi \hat H \Psi_u+2(J^{1/2}-X_u)\right]du.
\end{equation}
Here, the first term in the integrand corresponds to the kinetic energy of the system and the second term is the potential energy of surface capillary waves. In the
case of an infinitesimal amplitude of surface waves and the absence of energy pumping and dissipation,(\ref{eq1}) and (\ref{eq2}) are transformed to the dispersion relation (\ref{disp}). The corresponding quantities are transformed in the linear approximation as $ u\to x$, $Y\to \eta(x,t)$ and $\Psi\to \psi(x,t)$.

{\bf Simulation results.} The aim of this work is to numerically solve the system of nonlinear integrodifferential equations (\ref{eq1}) and (\ref{eq2}). The differential and integral operators are calculated using spectral methods with the total number of Fourier harmonics $N$, i.e., boundary conditions are periodic. The integration in time is performed by an explicit fourth order Runge–Kutta method with the step $dt$. All calculations are carried out in the periodic region with the length $L=2\pi$ with the following parameters: $dt=2.5\cdot 10^{-6}$, $N=8192$, $\gamma_0=10^{ -6}$, $k_d=750$, $F_0=10^5/N$, $k_1=3$, $k_2=7$. To suppress the aliasing effect, a low-frequency filter eliminating higher harmonics $k\geq N/3$ at each step of the integration in time.

Figure 1 shows the evolution of the total energy of the system given by Eq.(\protect\ref{ham2})at the calculation parameters presented above. It is seen that the energy under the action of the external chaotic force increases to a certain value and then undergoes rather complex oscillations near this value. The average energy in the quasistationary motion regime (wave turbulence) was $\langle H \rangle_t \approx 1$. The probability density function for the surface amplitude measured in the steady state is shown in the inset of Fig. 1. It is seen that the probability density function is very close to the Gaussian distribution shown by the red dashed line. This indicates the developed character of the observed capillary turbulence on the liquid surface.

\begin{figure}[t]
\centering
\includegraphics[width=.9\linewidth]{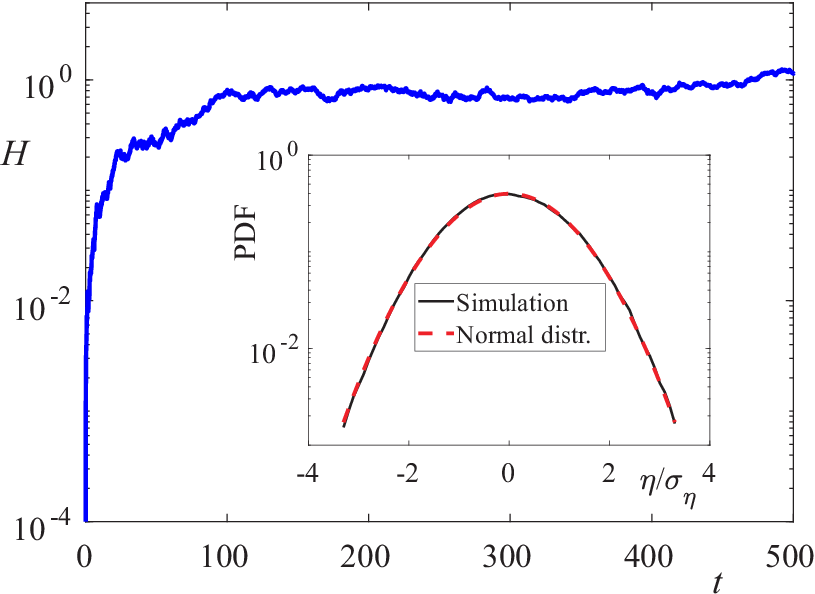}
\caption{(Color online) Time dependence of the total energy of the system given by Eq.(\ref{ham2}). The inset shows the probability distribution function (PDF) for surface amplitudes measured with respect to the standard deviation $\sigma_{\eta}$ in comparison with the Gaussian distribution presented by the red dashed line.}
\label{fig1}
\end{figure}

The motion of the liquid in the quasistationary state has a complex chaotic character (see Fig. 2). Figure 2 presents the shape of the liquid surface in the
quasistationary state at the time $t=410$. It is seen that the surface takes a complex irregular shape. The time averaged steepness of waves in the quasistationary
state is estimated as
$$\epsilon=\left\langle \sqrt{\frac{1}{L}\int|\eta(x,t)_x|^2dx}\right\rangle_t \simeq 0.3,$$
which is noticeably smaller than unity. Strongly nonlinear structures such as jets, droplets, and bubbles are not observed on the liquid surface at this steepness \cite{strong}. Thus, the external force (energy pumping) can transfer the system of nonlinear collinear capillary waves to the quasistationary chaotic state.

\begin{figure}[t]
\centering
\includegraphics[width=.9\linewidth]{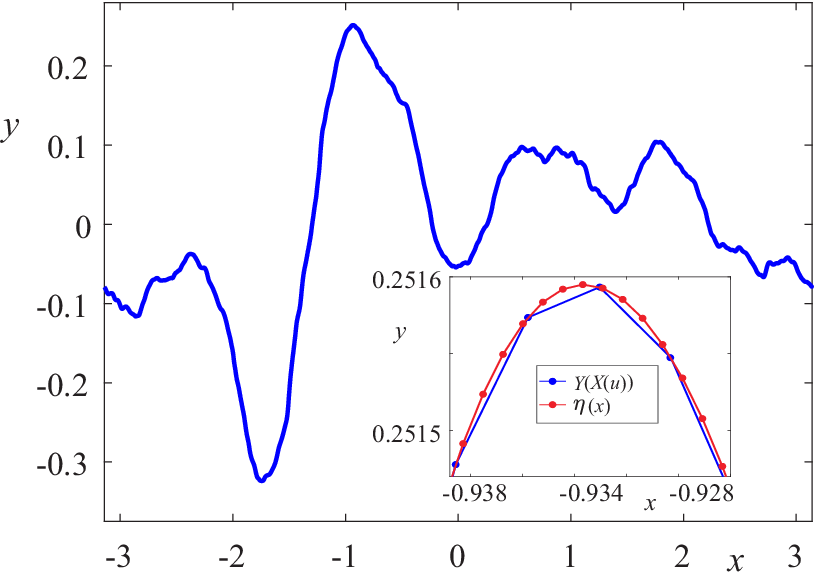}
\caption{Color online) Shape of the liquid surface in the quasistationary state at the time $t=410$. The inset shows the liquid surface in the region $x\in[-0.938, -0.928]$, the blue line corresponds to the function $Y(X)$ calculated from (\ref{eq1}) and (\ref{eq2}) and the red line is the result of interpolation on the uniform grid.}
\label{fig2}
\end{figure}

The main subject of this study is the turbulence spectrum of the system at the developed stage of evolution. The turbulence spectra given by Eqs. (1) and
(3) can in principle be applicable for the function $Y(u,t)$, calculated from Eqs. (\ref{eq1}) and (\ref{eq2}) because the dispersion relation specified by Eq. (7) is also valid for it. At the same time, the dependence $Y=Y(u,t)$ itself does not have any direct physical meaning because it describes the dynamics of perturbations of the coordinates $\{X,Y\}$ in the conformal space. For the total description of turbulence, it is necessary to explicitly express the shape of the surface $\eta(x,t)$ on a uniform grid of the horizontal $x$ axis. The parametric dependence $Y(X_i)$ obtained from Eqs (\ref{eq1}) and (\ref{eq2})is defined at sites $i$ of the calculated nonuniform grid $X_i(u,t)$. To obtain the explicit dependence $y=\eta(x,t)$, it is necessary to interpolate the values of the function $Y(X(u))$ at new sites $j$ with a fixed step $dx=X_{j+1}-X_{j}$. Cubic splines are used for the interpolation procedure. To the best of our knowledge, this method has not yet
been applied to analyze calculations based on dynamic conformal transforms. The result of interpolation is shown in the inset of Fig. 2. It is seen that the procedure reconstructs the profile with a high accuracy. The average error in the estimated potential energy of capillary waves by Eqs.~(\ref{Ham}) and (\ref{ham2}) is about $10^{-4}$. Such a high accuracy is reached due to a high density of the grid and to the absence of strongly nonlinear structures at which the function $Y(X(u))$ no longer a bijection.

\begin{figure}[h]
\centering
\includegraphics[width=0.9\linewidth]{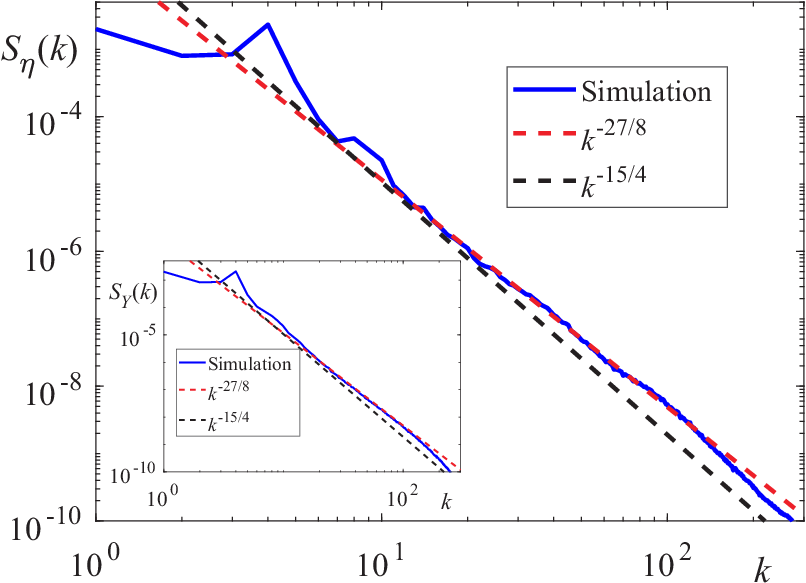}
\caption{(Color online) Time-averaged spatial spectrum of surface perturbations. The black dashed line is the Zakharov–Filonenko spectrum (\ref{ZF}), and the red dashed line is the turbulence spectrum given by Eq. (\ref{5w}). The inset shows the spatial spectrum of the function $Y(u,t)$.}
\label{fig3}
\end{figure}

Figure~\ref{fig3} shows the time-averaged spatial spectrum of the surface $S_{\eta}(k)=\left\langle |\eta_k|^2\right\rangle_t$ in the quasistationary motion regime. It is seen that the spectrum of surface perturbations has a power-law form in the wavenumber interval $k\in[10, 150]$. The exponent of the spectrum is noticeably different from that for the classical Zakharov–Filonenko spectrum~(\ref{ZF})obtained under the assumption of the dominant role of three-wave resonant interactions in the isotropic geometry. At the same time, the estimate given by Eq. (\ref{5w}) derived under the assumption of the dominant role of five-wave resonant interactions much better reproduces the numerical simulation data. The turbulence spectrum for the function $Y(u,t)$ is shown in the inset of Fig~\ref{fig3}. It is seen
that the spectrum $S_{Y}(k)=\left\langle |Y_k|^2\right\rangle_t$  is also in good agreement with Eq.(\ref{5w}).

\begin{figure}[t]
\centering
\includegraphics[width=0.9\linewidth]{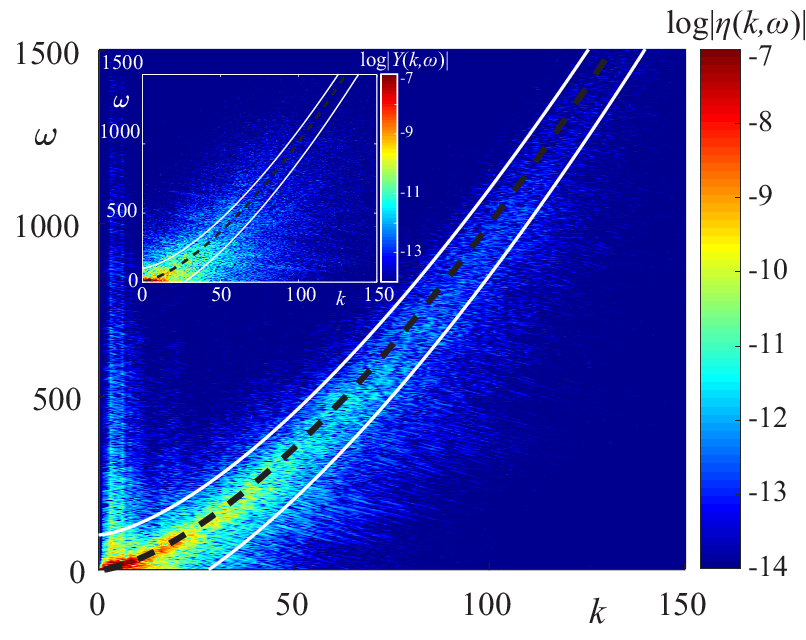}
\caption{(Color online) Log–lin $(\omega,k)$ map of the space-time Fourier transform $|\eta(k,\omega)|$. The black dashed line corresponds to the exact dispersion relation given by (\ref{disp}), and the white solid lines describe the nonlinear frequency broadening. The inset shows the Fourier transform for the function $Y(u,t)$.}
\label{fig4}
\end{figure}

\begin{figure}[t]
\centering
\includegraphics[width=0.9\linewidth]{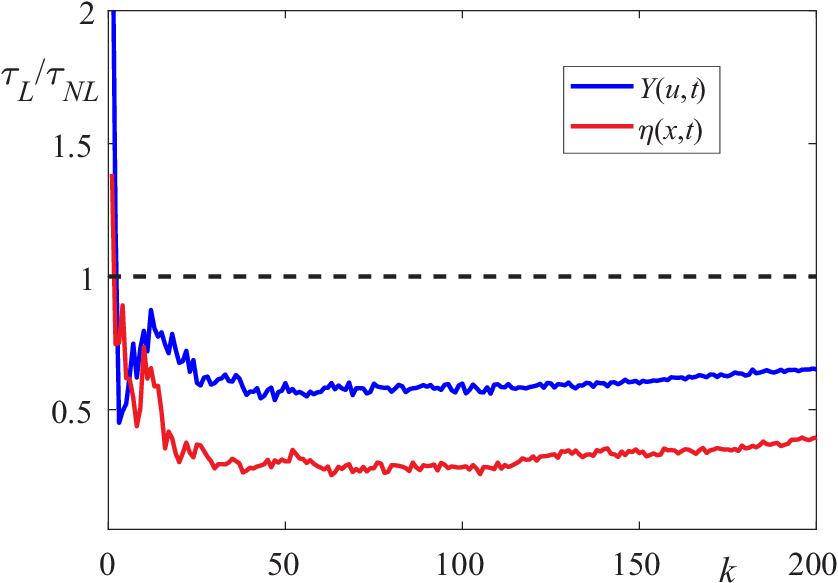}
\caption{(Color online) Linear-to-nonlinear time ratio $\tau_L/\tau_{NL}$ calculated for the functions (red line) $\eta(x,t)$ and (blue line) $Y(u,t)$.}
\label{fig5}
\end{figure}

To analyze the possible effect of coherent structures (e.g., solitons or bound waves \cite{bound}) on the evolution of the surface, the space–time Fourier transform
of the shape of the surface is plotted in Fig.~\ref{fig4}. It is seen that perturbations exist in the entire wavenumber range. Fourier harmonics are located in a narrow range along the dispersion relation (\ref{disp}). Any strongly nonlinear structures are not manifested. Figure~\ref{fig4} demonstrates only the broadening of frequencies due to nonlinear effects (amplitude dependence of the velocity of waves). It is remarkable that the broadening of frequencies is more noticeable for the spectrum $|Y(k,\omega)|$ shown in the inset of Fig.~\ref{fig4}. The average broadening of the frequency for the wavenumber $k$ can be calculated by the formula \cite{Naz22}:
\begin{equation}\nonumber
\delta_{\omega}(k)=\left[\frac{\int_0^{\infty}(\omega-\omega_k)^2|\eta(k,\omega)|^2 d\omega}{\int_0^{\infty}|\eta(k,\omega)|^2 d\omega}\right]^{1/2},
\end{equation}
where $\omega_k$ is determined from the dispersion relation (\ref{disp}). The parameter $\delta_{\omega}$ determines the characteristic nonlinear time $\tau_{NL}=1/\delta_{\omega}$, which is an important parameter in the theory of weak turbulence \cite{zakh}. The criterion of applicability of this theory has the form: $\tau_L/\tau_{NL}\ll 1$, where $\tau_L=1/\omega_k$ is the time determined from the linear dispersion relation. The calculated ratio $\tau_L/\tau_{NL}$  for the
functions $\eta$ and $Y$ is shown in Fig.~\ref{fig5}. It is seen that the broadening of frequencies is indeed larger for the function $Y(u,t)$. The ratio $\tau_L/\tau_{NL}$ for the reconstructed dependence is almost an order of magnitude smaller than unity. Thus, the comparison of characteristic times indicates a weakly nonlinear character of the evolution of the system.

{\bf Conclusion.}
To summarize, a new model has been proposed for the direct numerical simulation of wave turbulence appearing on the free surface of the liquid. The computational
model is fully nonlinear and involves effects of surface tension, as well as energy dissipation and pumping. Simulation results have shown that the system
of interacting nonlinear capillary waves can pass to the quasistationary state when the action of an external force is completely compensated by dissipative
effects. The motion of the liquid in this regime becomes complex and irregular, and the probability distribution function of the surface amplitude approaches the Gaussian distribution. The measured spectrum of surface perturbations in the quasistationary state takes a power-law form with the exponent close to that for the analytical spectrum obtained under the assumption of the dominant effect of five-wave resonant interactions in the anisotropic plane symmetric geometry. The analysis of the space–time Fourier transform also indicates a weakly nonlinear character of the evolution of waves. It is noteworthy that the numerical results are in good agreement with experimental studies with liquid mercury \cite{Ricard21}.


This work was supported by the Russian Science Foundation, project no. No. 19-72-30028.

\end{document}